\documentclass[11pt,a4paper]{article}
\usepackage{jheppub}

\makeatletter

\@addtoreset{equation}{section}
\makeatother

\newcommand{\red}{\textcolor{black}}

\title{\vbox{
\baselineskip 14pt
\hfill \hbox{\normalsize KUNS-2482}
} \vskip 1.7cm
\bf Flavor structure in  D-brane models: \\
Majorana neutrino masses\vskip 0.5cm
}

\author[]{Yuta Hamada,}
\author[]{Tatsuo Kobayashi}
\author[]{and Shohei Uemura}
\affiliation[]{Department of Physics, Kyoto University,\\ 
Kyoto 606-8502, Japan}
\emailAdd{hamada@gauge.scphys.kyoto-u.ac.jp}
\emailAdd{kobayashi@gauge.scphys.kyoto-u.ac.jp}
\emailAdd{uemura@gauge.scphys.kyoto-u.ac.jp}

\abstract{
We study the flavor structure in intersecting D-brane models.
We study anomalies of the discrete flavor symmetries.
We analyze the Majorana neutrino masses, which can be 
generated by D-brane instanton effects.
It is found that a certain pattern of mass matrix 
is obtained and the cyclic permutation symmetry remains unbroken.
As a result, trimaximal mixing matrix can be realized if Dirac neutrino mass and charged lepton mass matrices are diagonal.}
\arxivnumber{1402.2052}

\begin{document}
\maketitle

\renewcommand{\thefootnote}{\arabic{footnote}}
\setcounter{footnote}{0}

\section{Introduction}

The Standard Model has been confirmed by the discovery of 
the Higgs scalar and other precision measurements.
However, it has various mysteries still.
One of them is the mystery on the flavor structure.
Why are there three generations ?
Why are quark and lepton masses hierarchical ?
Which mechanism determines their mixing angles ?
Indeed, the Yukawa sector has most of free parameters  
in the Standard model.
Discrete flavor symmetries would be important 
to understand fermion masses and mixing angles 
\cite{Altarelli:2010gt,Ishimori:2010au,King:2013eh}.
For example, the mixing matrix in the lepton sector, 
the PMNS matrix, can be approximated by the tri-bimaximal mixing matrix 
in the limit $\theta_{13}=0$ \cite{Harrison:2002er}.
In field-theoretical model building, one starts with 
a large flavor symmetry.
Then, one assumes that the flavor symmetry breaks properly 
into $Z_3$ and $Z_2$ subsymmetries 
in the charged lepton or the neutrino masses, 
such that the tri-bimaximal mixing can be realized.

Superstring theory is a promising candidate for 
unified theory of all of the interactions including gravity 
and all of the matter fields and Higgs field(s) (see for a review \cite{Ibanez}).
It is found that superstring theory on six-dimensional 
compact space leads to interesting flavor structures.
In particular, certain types of four-dimensional 
superstring models with rather simple six-dimensional compact spaces 
such as tori and orbifolds lead to definite discrete flavor symmetries.
For example, intersecting D-brane models and magnetized D-brane models are 
among interesting model building in superstring theory 
\cite{Bachas:1995ik,Berkooz:1996km,Blumenhagen:2000wh,Aldazabal:2000dg,Angelantonj:2000hi,
Ibanez:2001nd} (see for review \cite{Blumenhagen:2006ci,Ibanez} and references therein).
These intersecting/magnetized D-brane models can lead to 
discrete flavor symmetries 
such as $D_4$, $\Delta(27)$, $\Delta(54)$ 
\cite{Abe:2009vi,Abe:2009uz,BerasaluceGonzalez:2012vb}.\footnote{See also \cite{Higaki:2005ie}.}
Similar discrete flavor symmetries can be derived in 
heterotic string theory on orbifolds \cite{Kobayashi:2004ya}.\footnote{
See for recent works on other discrete stringy symmetries, e.g.  
\cite{BerasaluceGonzalez:2011wy,Ibanez:2012wg,BerasaluceGonzalez:2012zn,
Anastasopoulos:2012zu,Honecker:2013hda,Bizet:2013gf,Nilles:2013lda,Bizet:2013wha} .}
In these models, we can calculate explicitly Yukawa couplings and higher order couplings 
\cite{Hamidi:1986vh,Cvetic:2003ch,Cremades:2004wa}

However, such discrete flavor symmetries may be 
broken by non-perturbative effects.
{}From such a viewpoint, anomalies of discrete symmetries 
\cite{Araki:2008ek,Araki:2007ss,Nilles:2013lda,Bizet:2013wha,Honecker:2013hda} 
are important because anomalous symmetries may be 
broken by non-perturbative effects.
Even anomaly-free U(1) gauge symmetries can be broken 
when axions couple with U(1) gauge bosons and they become massive.
Furthermore, as concrete non-perturbative effects, 
D-brane instanton effects have been studied \cite{Blumenhagen:2006xt} (see also for a review \cite{Blumenhagen:2009qh}
and references therein).
{}From the viewpoint of flavor physics, one of important points is that
D-brane instanton effects can generate 
right-handed Majorana neutrino masses \cite{Ibanez:2006da,Ibanez:2007rs,Cvetic:2007ku}.
Then, it is also important to investigate  patterns of right-handed 
Majorana neutrino mass matrices derived by D-brane instanton effects 
and study whether such effects break some or all of discrete flavor symmetries 
and which symmetries remain unbroken.

In this paper, we study the flavor structure in 
intersecting D-brane models as well as magnetized D-brane models.
We study anomalies of discrete flavor symmetries derived in 
intersecting D-brane models.
We also study right-handed Majorana neutrino mass matrices, 
which can be generated by D-brane instanton effects.
We show which types of Majorana mass matrices can be derived and 
which flavor symmetries remain unbroken even with 
right-handed Majorana neutrino mass matrices generated by 
D-brane instanton effects.

This paper is organized as follows.
In section 2, we review briefly the discrete flavor symmetries 
derived from intersecting D-brane models as well as 
magnetized D-brane models.
In section 3, we study anomalies of these discrete flavor 
symmetries\red{.}
In section 4, we study right-handed Majorana masses generated 
by D-brane instanton effects.
Section 5 is devoted to conclusion and discussion.
\red{In Appendix A, we show the computation to integrate non-vanishing
Wilson line phase.}

\section{Discrete flavor symmetries}

In this section, we review briefly discrete flavor symmetries 
appearing in intersecting D-brane models as well as magnetized D-brane models \cite{Abe:2009vi,BerasaluceGonzalez:2012vb}.
For concreteness, we consider IIA D6-brane models on $T^6=T^2_1\times T^2_2 \times T^2_3$, 
where each D6-brane wraps one-cycle of each $T^2$ of $T^6=T^2_1\times T^2_2 \times T^2_3$.
That is, our setup is as follows.
We consider $N_a$ stacks of D6-branes, which lead to $U(N_a)$ gauge symmetry, 
and they have winding numbers $(n_a^i,m_a^i)$ along the $x_i$ and $y_i$ directions 
on $T^2_i$, where 
we use orthogonal coordinates $(x_i,y_i)$ on $T^2_i$.
When we denote the basis of one-cycles on $T^2_i$ by $[a_i]$ and $[b_i]$, 
which correspond to  the $x_i$ and $y_i$ directions, 
the three-cycle, along which this set of D6-brane winds, is represented 
by 
\begin{eqnarray}
[\Pi_a] = \prod^3_{i=1} (n^i_a [a_i] + m^i_a[b_i]).
\end{eqnarray}

Here, we consider two sets of D-branes, one set is $N_a$ stacks of D6-branes 
and another is $N_b$ stacks of D6-branes.
These lead to $U(N_a)\times U(N_b)$ gauge groups.
Suppose that these two stacks of D6-branes intersect each other on $T^2_i$.
Their intersecting number on $T^2_i$ is obtained by 
\begin{eqnarray}
I_{ab}^{(i)}= (n^i_a m_b^i - m_a^in_b^i),
\end{eqnarray}
and their total intersecting number on $T^6$ is obtained by 
\begin{eqnarray}
[\Pi_a]\cdot[\Pi_b] = I_{ab} = \prod_{i=1}^3 I_{ab}^{(i)}.
\end{eqnarray} 
Then, chiral matter fields with bi-fundamental representations $(N_a,\bar N_b)_{(1,-1)}$ 
under $U(N_a)\times U(N_b)$ appear at intersecting points on $T^2_i$, 
where the index $(1,-1)$ denotes $U(1)^2$ charges inside $U(N_a)$ and $U(N_b)$.
There appear $I_{ab}$ families of bi-fundamental matter fields.
When $I_{ab}$ is negative, there appear $|I_{ab}|$ families of matter fields 
with the conjugate representation $(\bar N_a, N_b)_{(-1,1)}$.

The total flavor symmetry is a direct product of flavor symmetries appearing 
on one of $T^2_i$.
Thus, we concentrate on the flavor symmetry realized on one of $T^2_i$.
Then, we denote $I_{ab}^{(i)}=g$.
Theses modes on $T^2_i$ have definite $Z_g$ charges and $Z_g$ 
transformation is represented by 
\begin{eqnarray}
Z = \left(
\begin{array}{ccccc}
1 &  & &  & \\
 & \rho & & & \\
 & & \rho^2 & & \\
 & & & \ddots & \\
 & & & & \rho^{g-1}
\end{array}
\right),
\end{eqnarray}
where $\rho = e^{2\pi i/g}$.
In addition, there is a cyclic permutation symmetry $Z_g^{(C)}$ among 
these modes, i.e.
\begin{eqnarray}
C= \left(
\begin{array}{cccccc}
0 & 1 & 0 & 0 & \cdots & 0 \\
0 & 0 & 1 & 0 & \cdots & 0 \\
 & &     &   & \ddots & \\
 1 & & &    & \cdots & 0
\end{array}
\right).
\end{eqnarray}
Furthermore, these elements do not 
commute each other, 
\begin{eqnarray}
CZ = \rho Z C.
\end{eqnarray}
Thus, this flavor symmetry includes another $Z_g'$ symmetry, which 
is represented by 
\begin{eqnarray}
Z' = \left(
\begin{array}{ccccc}
\rho &  & &  & \\
 & & & \ddots & \\
 & & & & \rho
\end{array}
\right).
\end{eqnarray}
Then, these would generate the non-Abelian flavor symmetry, 
$(Z_g \times Z_g') \rtimes Z_g^{(C)}$.

For example, when $g=2$ and $g=3$, the symmetries correspond to 
$D_4$ and $\Delta(27)$.
In addition, when the totally D-brane system has the $Z_2$ reflection symmetry $P$ 
between $i$-th mode and $(g-i)$-th mode
the $\Delta(27)$ symmetry for $g=3$ is enhanced into $\Delta(54)$ \cite{Abe:2009vi}.

Similarly, we can discuss models with more than two sets of D-branes.
For example, suppose that we add $N_c$ stacks of D-branes to the above system, 
and that their intersecting numbers satisfy $G.C.D.(I_{ab}^{(i)},I_{ac}^{(i)},I_{bc}^{(i)})=d$.
Then, this model has the discrete flavor symmetry $(Z_d \times Z_d') \rtimes Z_d^{(C)}$.

The above result is applicable to intersecting D-brane models on orientifolds 
through simple extension.
Also, we can extend our discussions to orbifold cases \cite{Abe:2009vi,Abe:2009uz,BerasaluceGonzalez:2012vb}.

Since magnetized D-brane models are T-duals to intersecting D-brane models, 
the magnetized D-brane models also have the same discrete flavor symmetries.
For example, we start with $(N_a + N_b)$ stacks of D9-branes on $T^6$.
Then, we introduce the magnetic flux on $T_i^2$ along $U(1)_a$ and $U(1)_b$ directions in
$U(N_a + N_b)$ as 
\begin{eqnarray}
F^{(i)} = 2\pi \left(
\begin{array}{cccccc}
M_a^{(i)} & & & & & \\
    & \ddots & & & & \\
    & & M_a^{(i)} & & & \\
    & & & M_b^{(i)} & &  \\
    & & &  & \ddots &  \\
    & & & & & M_b^{(i)} \\
    \end{array} \right),
    \end{eqnarray}
where $M_a^{(i)}$ and $M_b^{(i)}$ are integers.
This magnetic flux background breaks the gauge group 
$U(N_a + N_b)$ into $U(N_a) \times U(N_b)$.
The gaugino fields in the off-diagonal part correspond 
to the $(N_a,\bar N_b)$ bi-fundamental matter fields 
under the unbroken $U(N_a) \times U(N_b)$ gauge symmetry.
Zero-modes with such representation appear in this model, 
and the number of zero-modes on $T^2_i$ is equal to 
$M_a^{(i)}-M_b^{(i)}$.
When we denote $M_a^{(i)}-M_b^{(i)}=g$, this magnetized 
D-brane model leads to the same discrete flavor symmetry, 
$(Z_g \times Z_g')\rtimes Z_g^{(C)}$ as 
the above intersecting D-brane model.

\section{Discrete anomalies}

In this section, we study anomalies of discrete flavor symmetries.

\subsection{$U(1)$ anomalies}

Before studying anomalies of discrete flavor symmetries, 
it is useful to review anomalies of $U(1)$ gauge symmetries.
In this subsection, we give a brief review on $U(1)$ anomalies \cite{Aldazabal:2000dg,Cvetic:2001nr} 
(see also \cite{Blumenhagen:2006ci,Ibanez}).

First of all, we consider the torus compactification.
A D6-brane has a charge of RR 7-form $C_7$.
The total charge should vanish in a compact space.
That leads to the following tadpole cancellation condition,
\begin{eqnarray}
\sum_a N_a [\Pi_a] =0.
\end{eqnarray}

The $SU(N_a)^3$ anomaly coefficient is calculated in the 
intersecting D-brane models by 
\begin{eqnarray}
A_a = \sum_b I_{ab} N_b,
\end{eqnarray}
because there are $I_{ab}$ matter fields with $(N_a,\bar N_b)_{(1,-1)}$ 
for $I_{ab} >0$ and  $|I_{ab}|$ matter fields with $(\bar N_a,N_b)_{(-1,1)}$ 
for $I_{ab} <0$.
However, the tadpole cancellation condition leads 
to
\begin{eqnarray}
[\Pi_a]\cdot  \sum_b N_b [\Pi_b] = 0.
\end{eqnarray}
That implies that $A_a = 0$, that is, 
anomaly free.

The $U(1)_a \times SU(N_b)^2$ mixed anomaly 
coefficient is obtained by 
\begin{eqnarray}
A_{ab} = N_a I_{ab}.
\end{eqnarray}
This anomaly is not always vanishing.
However, this anomaly can always be canceled 
by the Green-Schwarz mechanism, where 
an axion shifts under the $U(1)$ gauge transformation 
and the anomalous U(1) gauge boson becomes massive.

The  $U(1)$-gravity$^2$ anomaly coefficient is 
obtained by 
\begin{eqnarray}
A_{a-{\rm grav}} = N_a \sum_b I_{ab}N_b.
\end{eqnarray}
This anomaly is always vanishing when the tadpole cancellation 
condition is satisfied.

Next, we review on anomalies for the orientifold compactification.
That is, we introduce $O6$-branes along the direction 
$\prod_i [a_i]$.
The system must be symmetric under the $Z_2$ reflection, $y_i \rightarrow -y_i$.
In this case, we have to introduce a mirror $D6_{a'}$-branes 
with the winding number $(n_a^i,-m_a^i)$ corresponding to  $(n_a^i,m_a^i)$.
The O6-brane has (--4) times as RR charge as a D6-brane.
Then, the RR-tadpole cancellation condition requires 
\begin{eqnarray}
\sum_a N_a([\Pi_a] + [\Pi_{a'}])  - 4 [\Pi_{O6}] =0.
\end{eqnarray}
\begin{eqnarray}
\sum_{a \neq b}N_a [\Pi_b]\cdot ([\Pi_a] + [\Pi_{a'}]) 
+ N_b [\Pi_b]\cdot [\Pi_{b'}]  - 4[\Pi_b]\cdot [\Pi_{O6}]  =0.
\end{eqnarray}

In addition to $I_{ab}$ families of $(N_a,\bar N_b)_{(1,-1)}$ 
matter fields, there appear $I_{ab'}$ families of 
$(N_a,N_b)_{(1,1)}$ matter fields.
Moreover, there appear matter fields with symmetric 
and asymmetric representations under $U(N_a)$ with 
charge 2.
Their numbers are obtained by 
\begin{eqnarray}
& & \#_{a, {\rm asymm}} = \frac12 ([\Pi_a] \cdot [\Pi_{a'}]-
[\Pi_a]\cdot [\Pi_{O6}]) +[\Pi_a]\cdot [\Pi_{O6}], \\
& & \#_{a, {\rm symm}} = \frac12 ([\Pi_a] \cdot [\Pi_{a'}]-
[\Pi_a]\cdot [\Pi_{O6}]).
\end{eqnarray}
In this case, we can show that 
the $SU(N_a)^3$ anomaly coefficient always vanishes 
when the RR-tadpole cancellation condition is satisfied,  
similarly to in the torus compactification.
Also, the $U(1)_a-SU(N_b)^2$ anomaly coefficient 
is not always vanishing, but such anomaly can be 
canceled by the Green-Schwarz mechanism.

Finally, the $U(1)_a-$gravity$^2$ anomaly coefficient 
is obtained by 
\begin{eqnarray}
A_{a-{\rm grav}} &=& \prod_{b \neq} N_a N_b ([\Pi_{a}]\cdot [\Pi_b] + 
[\Pi_a]\cdot [\Pi_{b'}])  +2 \frac{N_a(N_a-1)}{2} \#_{a ,{\rm asymm}} \nonumber \\
& & 
+2 \frac{N_a(N_a+1)}{2} \#_{a ,{\rm symm}} \nonumber \\
&=& 3N_a[\Pi_a]\cdot[\Pi_{O6}].
\end{eqnarray}
This does not always vanish, but such anomaly can be 
canceled by the Green-Schwarz mechanism.

\subsection{Discrete anomalies}

In the gauge theory with the gauge group $G$ 
and the Abelian discrete symmetry $Z_N$, 
the $Z_N-G^2$ mixed anomaly  coefficient 
is calculated by \cite{Ibanez:1991hv,Banks:1991xj,Araki:2008ek,Ishimori:2010au}, 
\begin{eqnarray}
A_{Z_N-G^2}  =\sum_m q^{(m)} T_2({\bf R}^{(m)}),
\end{eqnarray}
where the summation of $m$ is taken over 
fermions with $Z_N$ charges $q^{(m)}$ and 
the representation ${\bf R}^{(m)}$ under $G$. 
Here, $T_2({\bf R}^{(m)})$ denotes 
the Dynkin index and we use the normalization such that $T_2=1/2$ for the fundamental 
representation of $SU(N)$.
When the following condition is satisfied \cite{Ibanez:1991hv,Banks:1991xj,Araki:2008ek,Ishimori:2010au}, 
\begin{eqnarray}
\sum_m q^{(m)} T_2({\bf R}^{(m)}) =0 ~~({\rm mod}~~N/2),
\end{eqnarray}
the $Z_N$ symmetry is anomaly-free.
Similarly, we can calculate the $Z_N$-gravity$^2$ anomaly 
coefficient by Tr$ q^{(m)}$.
If Tr$ q^{(m)} = 0$ (mod $N/2$), $Z_N$ is anomaly-free.
For example, $Z_2$ symmetry is always anomaly-free.

Each generator of non-Abelian discrete symmetries 
corresponds to an Abelian symmetry.
Thus, if each Abelian generator of non-Abelian 
discrete flavor symmetry satisfies the above 
anomaly-free condition, the total non-Abelian 
symmetry is anomaly-free.
When some discrete Abelian symmetries are anomalous, 
the total non-Abelian discrete symmetry is broken, 
and the subgroup, which does not include anomalous generators, 
remains unbroken.

In the non-Abelian discrete symmetry, 
there appear multiplets and each generator is  
represented by a matrix, $M$.
When $\det M=1$, the corresponding Abelian 
discrete symmetry is always anomaly-free.
Only multiplets with $\det M \neq 1$ can contribute 
on anomalies.
Since we have $\det Z' = 1$, 
the corresponding $Z_g'$ symmetry is always anomaly-free.
On the other hand, we find $\det Z = \det C= 1$  for $g=$ odd 
and $\det Z = \det C= -1$  for $g=$ even.
That means that the discrete flavor symmetry 
$(Z_g \times Z_g') \rtimes Z_g^{(C)}$ is always anomaly-free for $g=$ odd, 
but 
$Z_g$ and $Z_g^{(C)}$ can be anomalous 
for $g=$ even.
In particular, their $Z_2$ parts are anomalous.
One has to check the anomaly-free condition for 
such $Z_2$ part for $Z_g$ and $Z_g^{(C)}$.
For example, the $\Delta(27)$ flavor symmetry for $g=3$ is always anomaly-free.
However, $Z_2$ subgroups of $D_4$  for $g=2$ corresponding to the 
following elements,
\begin{eqnarray}
\left(
\begin{array} {cc}
1 & 0  \\
0 & -1
\end{array}
\right),\qquad 
\left(
\begin{array} {cc}
0 & 1  \\
1 & 0
\end{array}
\right),
\end{eqnarray} 
can be anomalous.

First, we discuss the torus compactification.
For simplicity, we concentrate on the flavor symmetry 
appearing the first torus $T^2_1$ and we assume that 
all of intersecting numbers on $T^2_1$, $I_{ab}^{1}$, are 
even.
Thus, the total flavor symmetry includes the 
$Z_2$ symmetry as well as $Z_2^{(C)}$, which can be anomalous.
Also, we assume that there appears a trivial symmetry from 
the other $T^2_2 \times T^2_3$.
Now, let us examine the $Z_2-SU(N_a)^2$ anomaly.
There are $I_{ab}$ bi-fundamental matter fields 
with the representation $(N_a,\bar N_b)$.
A half of $I_{ab}$ matter fields have even $Z_2$ charge and 
the others have odd $Z_2$ charge.
The anomaly coefficient of $Z_2-SU(N_a)^2$ anomaly can be 
written by
\begin{eqnarray}
\sum_b \frac{I_{ab}}{2} N_b \frac12.
\end{eqnarray}
It vanishes because the tadpole cancellation condition, 
$\sum_b I_{ab}N_b=0$.
Thus, this $Z_2$ symmetry is anomaly-free on the torus compactification.
Since only this $Z_2$ symmetry can be anomalous and the others are 
always anomaly-free, 
the non-Abelian flavor symmetries $(Z_g \times Z_g') \rtimes Z_g^{(C)}$ 
are always anomaly-free in the torus compactification.

Next, we study the orientifold compactification.
Similarly, we can calculate  the $Z_2-SU(N_a)^2$ anomaly coefficient,
\begin{eqnarray}
& & \sum_{b \neq a} \left( \frac{I_{ab}}{2} N_b  + \frac{I_{ab'}}{2} N_{b'}\right) \frac12 
+ \frac{N_a-2}{4} \#_{a, {\rm asymm}}  
+ \frac{N_a+2}{4} \#_{a, {\rm symm}} \nonumber \\
&=& {\frac{[\Pi_a]\cdot[\Pi_{O6}]}{2}}
\end{eqnarray}
That is not always vanishing, but it is proportional to the 
$U(1)_a$-grav$^2$ anomaly.
Thus, this anomaly could be canceled when one requires the 
axion shift under the $Z_2$ transformation, which is related with 
the axion shift under $U(1)_a$.
In addition, when D6$_a$ branes are parallel to the O6-branes, 
$Z_2-SU(N_a)^2$ anomaly coefficient is always vanishing.

\section{Majorana neutrino masses}

In the previous section, we have studied on 
anomalies of discrete flavor symmetries.
Certain symmetries are anomaly-free.
For example, the $\Delta(27)$ flavor symmetry is anomaly-free.
Anomalous symmetries can be broken by non-perturbative effects.
There is no guarantee that anomaly-free 
symmetries are not broken by stringy non-perturbative effects.
In this section, we consider D-brane instanton effects as 
concrete non-perturbative effects.
We study which form of right-handed Majorana neutrino mass matrix can 
be generated by D-brane instanton effects.
Indeed, following \cite{Blumenhagen:2006xt,Blumenhagen:2009qh,Ibanez:2006da}, 
we study the sneutrino mass matrix assuming that 
the neutrino mass matrix has the same form and supersymmetry breaking 
effects are small.

\subsection{Neutrino mass matrix}

Here, we study right-handed Majorana neutrino masses, 
which can be generated by D-brane instanton effects.
We assume that $g$ families of right-handed neutrinos $\nu_R^a$
appear by intersections between D6$_c$-brane 
and D6$_d$ branes, and that  
 their intersecting numbers are equal to 
$I_{cd}^{(i)} = g$ for the i-th $T^2$ and 
$I_{cd}^{(j)} =1$ for the other tori.
For the moment, let us concentrate on the three-generation model, 
$I_{cd}=3$, which can be obtained by 
$(I^{(1)}_{cd},I^{(2)}_{cd},I^{(3)}_{cd}) = (\underline{3,1,1})$, 
where the underline denotes all the possible permutations. 
We consider D2-brane instanton, which wraps 
one-cycle of each $T^2$ of $T^6=T^2\times T^2\times T^2$.
We call it $D2_M$-brane.
It intersects with D6$_{c}$ brane and D6$_{d}$ brane.
At these intersecting points, zero-modes $\alpha_i$ 
and $\gamma_j$ appear and their numbers are 
obtained by $I_{Mc}$ and $I_{dM}$.
Only if there are two zero-modes for both 
$\alpha_i$ and $\gamma_j$ the neutrino masses can be 
generated by D2-brane instanton effect \cite{Blumenhagen:2006xt,Blumenhagen:2009qh,Ibanez:2006da}, 
\begin{eqnarray}
& & M\int d^2\alpha d^2 \gamma ~e^{-d^{ij}_a \alpha_i \nu_R^a \gamma_j} 
= Mc_{ab}, \nonumber \\
& & c_{ab} = 
\nu^a_R \nu^b_R (\varepsilon_{ij} \varepsilon_{k \ell}d^{ik}_a d^{j \ell}_b),
\end{eqnarray}
where the mass scale $M$ would be determined by the string scale $M_{st}$ and the 
instanton world volume $V$ as $M = M_{st}e^{-V}$.  
Here, $d^{ij}_a$ is the 3-point coupling coefficient among 
$\alpha_i$, $\nu_R^a$ and $\gamma_j$ \cite{Cvetic:2003ch}, which we show explicitly 
in the next subsection.
The 3-point coupling coefficient $d_a^{ij}$ can be written by 
$d_a^{ij}=d_{a1}^{ij}d_{a2}^{ij}d_{a3}^{ij}$, where 
$d_{ak}^{ij}$ for $k=1,2,3$ is the contribution from the $k$-th torus.
In addition, when $\alpha_i$, $\gamma_j$, or $\nu^a$ are localized at a single 
intersecting point on the $k$-th torus, we omit the indexes such as 
$d_{ak}^{j}$, $d_{ak}^{i}$, or $d_{k}^{ij}$.

We have to take into account  all of the possible $D2_M$-brane configurations, 
which can generate the above neutrino mass terms.
One can obtain two zero-modes of $\alpha_i$ and $\gamma_j$ 
for the $D2_M$-brane set corresponding to Sp(2) or U(2) gauge group 
with the intersecting numbers $|I_{Mc}|=|I_{Md}|=1$ \cite{Ibanez:2007rs} 
or a single $D2_M$-brane with the intersecting numbers, $|I_{Mc}|=|I_{Md}|=2$.

When the $D2_M$-brane set corresponds to the Sp(2) or U(2) brane, 
the zero-modes, $\alpha_i$ and $\gamma_j$, are doublets and 
the gauge invariance allows the certain couplings, say $\alpha_i$ and $\gamma_i$, 
but not $\alpha_i$ and $\gamma_j$ for $i \neq j$.
When $I_{Mc}=I_{dM}=1$, the following form of the Majorana mass is generated, 
\begin{eqnarray}
\int d^2\alpha d^2 \gamma ~e^{-d^{11}_a \alpha_1 \nu_R^a \gamma_1-
d^{22}_a \alpha_2 \nu_R^a \gamma_2} 
= \nu^a_R \nu^b_R d^{11}_a d^{22}_b.
\end{eqnarray}
More explicitly, the following form of mass matrix is obtained \cite{Ibanez:2007rs}, 
\begin{eqnarray}
M c_{ab} = \left(
\begin{array}{ccc}
d_{1}^{11} d_{1}^{22} & d_{1}^{11}d_{2}^{22} &d_{1}^{11}d_{3}^{22} \\
d_{2}^{11} d_{1}^{22} & d_{2}^{11}d_{2}^{22} &d_{2}^{11}d_{3}^{22} \\
d_{3}^{11} d_{1}^{22} & d_{3}^{11}d_{2}^{22} &d_{3}^{11}d_{3}^{22} \\
\end{array}
\right).
\end{eqnarray}
This Majorana mass matrix has the rank one.
However, we have to take into account all of the $D2_M$-brane configurations, 
that is, the position of $D2_M$-brane sets.
Thus, we integrate over the position of the $D2_M$-brane sets.
Such integration over the $D2_M$-brane position would recover the 
cyclic permutation symmetry, $Z_{g=3}^{(C)}$.
Then, we would obtain the following form of Majorana neutrino mass 
matrix,
\begin{eqnarray}\label{eq:n-mass-33}
M = \left(
\begin{array}{ccc}
A & B & B\\
B & A & B\\
B & B & A\\
\end{array}
\right).
\end{eqnarray}
We will show this form by an explicit calculation 
in the next subsection.
As a result, there remains the cyclic permutation symmetry, $Z_{g=3}^{(C)}$, 
unbroken,
 but $Z_{g=3}$ and $Z_{g=3}'$ symmetries are broken 
by D-brane instanton effects, which generate 
the Majorana neutrino masses.
This form also has the $Z_2$ reflection symmetry $P$.
Thus, if the full D-brane system has the $Z_2$ reflection symmetry, 
the symmetry is enhanced into $S_3$.

Similarly, we can study a single $D2_M$-brane with 
the intersecting numbers, $|I_{Mc}|=|I_{Md}|=2$.
There are two types of $D2_M$-brane instanton configurations 
leading to $|I_{Mc}|=|I_{Md}|=2$.
In one type, we have the configuration with 
$|I_{Mc}^{(j)}| = |I_{Md}^{(k)}| =2$ 
for $j \neq k$, and in the other type we have 
the configuration with $|I_{Mc}^{(j)}| = |I_{Md}^{(j)}| =2$.

In the first case with $|I_{Mc}^{(j)}| = |I_{Md}^{(k)}| =2$ 
for $j \neq k$, let us set e.g. $j=1$ and $k=2$.
Then, the Yukawa coupling $d^{ij}_a$ can be written by 
$d^{ij}_a=d^i_{a1}d^j_{a2}d_{a3}$.
Also we assume that $I_{cd}^{(1)}=3$ and $I_{cd}^{(2)}=I_{cd}^{(3)}=1$.
Then, the neutrino mass can be written by 
\begin{equation}
 \varepsilon_{ij} \varepsilon_{k\ell} d_{a}^{ik} d_{b}^{jl}=\varepsilon_{ij} \varepsilon_{kl} d_{a1}^{i}d_{2}^k 
d_{3} d_{b1}^{j}d_{2}^\ell d_{3}. 
\end{equation}
However, this vanishes identically \cite{Ibanez:2006da}.
We obtain the same result for $|I_{Mc}^{(j)} |= |I_{Md}^{(k)} |=2$ with $j \neq k$, 
when $(I^{(1)}_{cd},I^{(2)}_{cd},I^{(3)}_{cd}) = (\underline{3,1,1})$.

On the other hand, if a single $D2_M$-brane configuration with 
$|I_{Mc}^{(j)}| = |I_{Md}^{(j)}| =2$ is possible, we obtain the non-vanishing 
neutrino mass matrix $Mc_{ab}$.
Then, when we integrate over the position of the $D2_M$-brane instanton, 
we would obtain the same results as Eq.(\ref{eq:n-mass-33}).
Thus, the cyclic permutation symmetry $Z_{g=3}^{(C)}$ is recovered.

This result can be extended for models with $g$ flavors of neutrinos.
When we take into account all of the possible D-brane instanton configurations, 
we would realize the neutrino mass matrix $Mc_{ab}$ with the cyclic permutation symmetry 
$Z_g^{(C)}$, i.e.
\begin{eqnarray}
c_{ab}=c_{a'b'}   {\rm~~~for~~~}a'=a+1,~b'=b+1.
\end{eqnarray}
Also the mass matrix is symmetric, i.e. $c_{ab} = c_{ba}$.
For example, we obtain 
\begin{eqnarray}
c_{ab}=\left(
\begin{array}{cc}
A & B \\
B & A 
\end{array}
\right),
\end{eqnarray}
for $g=2$ and 
\begin{eqnarray}
c_{ab}=\left(
\begin{array}{cccc}
A & B & B' & B\\
B & A & B & B'\\
B' & B & A & B \\
B & B' & B & A 
\end{array}
\right),
\end{eqnarray}
for $g=4$. 
It is found that the D-brane instantons break $Z_g'$ into $Z_2$ if 
$g$ is even.
Otherwise, the $Z_g'$ symmetry as well as the $Z_g$ 
symmetry is  completely broken.
However, the cyclic permutation symmetry remains.\footnote{
These forms also have the $Z_2$ reflection symmetry.}

We have studied the neutrino mass matrix 
by assuming that the neutrino and sneutrino have 
the same mass matrix and supersymmetry breaking effect is small 
\cite{Blumenhagen:2006xt,Blumenhagen:2009qh,Ibanez:2006da}.
The important point to derive our result is the cyclic permutation symmetry.
Thus, we would obtain the same result if the D-brane instatons 
do not break such a symmetry but supersymmetry is broken.

\subsection{Explicit computation}

Here, we discuss the Majorana neutrino mass matrix by computing explicitly 
the three-generation models.
We consider the D2-brane instanton corresponding to 
 Sp(2) or U(2) gauge symmetry.
Suppose that D6$_c$ and D6$_d$ branes have the intersecting number,
$I_{cd}=3$, and at three intersecting points there appear 
three generations of right-handed neutrinos.
We set $(I^{(1)}_{cd},I^{(2)}_{cd},I^{3)}_{cd})=(3,1,1)$, 
and $I_{Mc}=I_{dM}=1$.
Because the right-handed neutrinos are localized at 
different points from each other on the first torus, 
only the first torus is important for the flavor symmetry.
Thus, we concentrate on the first torus 
for computations on Yukawa couplings and 
Majorana masses.
We also omit the index corresponding to the 
$k$-torus.
In the following computations, we set Wilson line moduli zero because it dose not affect the flavor structure
(\red{see} appendix A for more detail).

There are three generations of $\nu_a$ and we here label their flavor index as $a=0,1,2$.
Also there are two-zero modes, 
$\alpha_i$ and $\gamma_j$ $(i,j=1,2)$, but note that these indexes $i,j$ correspond to 
the doublets under $Sp(2)$ or $U(2)$ and the intersecting numbers, $I_{cM}$, and $I_{Md}$, 
are equal to one,  $I_{cM}=I_{Md}=1$.

Suppose that there are three fields $\phi_a$, $\chi_{i'}$ and $\chi_{j'}$ with 
the "flavor numbers", $a=0,\cdots,I_{cd}-1$, $i'=0,\cdots,I_{dM}-1$, and 
$j'=0,\cdots,I_{Mc}-1$, where 
$I_{cd}$, $I_{dM}$, and $I_{Mc}$ are the corresponding intersecting numbers on the torus.
In this case, the 3-point couplings  $d_a^{i'j'}$ among three fields 
can be calculated by \cite{Cvetic:2003ch}
\begin{equation}d_{a}^{i'j'}=C \sum_{\ell \in {Z}} {\rm exp}\left( \frac{-A_{ai'j'}(\ell)}{2\pi \alpha'} \right),
\end{equation}
where $C$ is a flavor-independent constant due to quantum contributions and 
\begin{equation} A_{ai'j'}(\ell)=\frac{1}{2} A |I_{cd} I_{dM} I_{Mc}|\left( \frac{a}{I_{cd}}
 + \frac{i'}{I_{dM}}+ \frac{j'}{I_{Mc}}+\frac{\varepsilon}{I_{dM}I_{Mc}} +\ell \right)^2,
\end{equation}
and $A$ denotes the area of the first torus.
Here, $\varepsilon$ denotes the position of $D2_M$-brane on the first torus and we normalize 
$\varepsilon$ such that $\varepsilon$  varies $[0,1]$ on the torus.
Note that this coupling corresponds to the contribution on the first torus, which determines 
the flavor structure, but we have omitted the index corresponding to the first torus.

By using the $\vartheta$-function, 
\begin{equation} \vartheta \left[
\begin{array}{c}
a \\
b \\
\end{array}\right] (\nu,\tau) = \sum_{\ell \in {Z}} {\rm exp} \left[ \pi i (a+\ell)^{2} \tau + 2 \pi i (a+\ell)(\nu + b) 
\right],
\end{equation}
we can write \begin{equation} d_{a}^{i'j'} = C \vartheta \left[
\begin{array}{c}
\frac{a}{I_{cd}} + \frac{i'}{I_{dM}} + \frac{j'}{I_{Mc}}+\frac{\varepsilon}{I_{dM} I_{Mc}} \\
0  \\
\end{array}
\right] \left( 0,\frac{iA\red{|}I_{cd} I_{dM} I_{Mc}|}{4\pi^{2}\alpha'} \right) .
\label{eq:3pcoupling}
\end{equation}

Our model corresponds to $a=0,1,2$, $I_{cd}=3$, $i'=j'=0$, $I_{dM}=I_{Mc}=1$.
In the above model, the 3-point couplings among 
$\nu_a$, $\alpha_i$, and $\gamma_j$ are written by
\begin{equation} d_{a}^{ij} = \delta_{ij} \vartheta \left[
\begin{array}{c}
-\frac{a}{3} + \varepsilon \\
0  \\
\end{array}
\right] \left( 0,\frac{3iA}{4\pi^{2}\alpha'} \right). 
\end{equation}
Recall again that the indexes $i$ and $j$ of $\alpha_i$ and $\gamma_j$ 
are doublet indexes under Sp(2) or U(2).

Using this, the matrix $c_{ab}$ is written by the integration of 
the position $\varepsilon$ over $[0,1]$,
\begin{eqnarray}c_{ab} & = & \int_{0}^{1} d\varepsilon \vartheta \left[
\begin{array}{c}
-\frac{a}{3} + \varepsilon \\
0  \\
\end{array}
\right] \left( 0,\frac{3iA}{4\pi^{2}\alpha'} \right)
\vartheta \left[
\begin{array}{c}
-\frac{b}{3} + \varepsilon \\
0  \\
\end{array}
\right] \left( 0,\frac{3iA}{4\pi^{2}\alpha'} \right) \nonumber \\
& = &
\int_{0}^{1} d\varepsilon \sum_{m=1}^{2} 
\vartheta \left[
\begin{array}{c}
-\frac{a}{6} - \frac{b}{6} + \varepsilon + \frac{m}{2}\\
0  \\
\end{array}
\right] \left( 0,\frac{3iA}{2\pi^{2}\alpha'} \right)  \\
& & \times \vartheta \left[
\begin{array}{c}
-\frac{a}{6} +\frac{b}{6} +\frac{m}{2} \\
0  \\
\end{array}
\right] \left( 0,\frac{3iA}{2\pi^{2}\alpha'} \right). \nonumber
\end{eqnarray}
We obtain 
\begin{eqnarray} 
&& 
\int_{0}^{1} d\varepsilon  
\vartheta \left[
\begin{array}{c}
-\frac{a}{3}  + \varepsilon + \frac{m}{2}\\
0  \\
\end{array}
\right] \left( 0,\frac{3iA}{2\pi^{2}\alpha'} \right) \nonumber \\
 & & = 
\int_{0}^{1} d\varepsilon  
\sum_{l \in {Z}} {\rm exp}\left[\pi i (-a/3 + \varepsilon +m/2 +\ell)^{2}\left(\frac{3 i A}{2 \pi^{2} \alpha'} \right) \right] \nonumber \\
& &= 
\int_{-\infty}^{\infty} dx {\rm exp} \left[-\frac{3A}{2\pi \alpha'} ( x - a/3 +m/2 )^{2} \right]\\
& & = 
\sqrt{\frac{2\pi^{2}\alpha'}{3A}}. \nonumber 
\end{eqnarray}
Using it, the matrix elements $c_{ab}$ can be computed as follows.
It is found that the diagonal elements $c_{aa}$ 
do not depend on $a$ and they are written by 
\begin{equation} c_{aa} = \sqrt{\frac{2\pi^{2}\alpha'}{3A}} 
\left(\vartheta \left[
\begin{array}{c}
0 \\
0  \\
\end{array}
\right] \left( 0,\frac{3iA}{2\pi^{2}\alpha'} \right)
+
\vartheta \left[
\begin{array}{c}
\frac{1}{2} \\
0  \\
\end{array}
\right] \left( 0,\frac{3iA}{2\pi^{2}\alpha'} \right)
\right).
\end{equation}
Similarly, the off-diagonal elements are written by
\begin{equation} c_{01} = \sqrt{\frac{2\pi^{2}\alpha'}{3A}} 
\left(\vartheta \left[
\begin{array}{c}
\frac{1}{6} \\
0  \\
\end{array}
\right] \left( 0,\frac{3iA}{2\pi^{2}\alpha'} \right)
+
\vartheta \left[
\begin{array}{c}
\frac{2}{3} \\
0  \\
\end{array}
\right] \left( 0,\frac{3iA}{2\pi^{2}\alpha'} \right)
\right),
\end{equation}
\begin{equation} c_{02} = \sqrt{\frac{2\pi^{2}\alpha'}{3A}} 
\left(\vartheta \left[
\begin{array}{c}
\frac{1}{3} \\
0  \\
\end{array}
\right] \left( 0,\frac{3iA}{2\pi^{2}\alpha'} \right)
+
\vartheta \left[
\begin{array}{c}
\frac{5}{6} \\
0  \\
\end{array}
\right] \left( 0,\frac{3iA}{2\pi^{2}\alpha'} \right)
\right),
\end{equation}
\begin{equation} c_{12} = \sqrt{\frac{2\pi^{2}\alpha'}{3A}} 
\left(\vartheta \left[
\begin{array}{c}
\frac{1}{6} \\
0  \\
\end{array}
\right] \left( 0,\frac{3iA}{2\pi^{2}\alpha'} \right)
+
\vartheta \left[
\begin{array}{c}
\frac{2}{3} \\
0  \\
\end{array}
\right] \left( 0,\frac{3iA}{2\pi^{2}\alpha'} \right)
\right).
\end{equation}

However, we have the following formula of the $\vartheta$-function
\begin{equation}\vartheta \left[
\begin{array}{c}
a \\
b  \\
\end{array}
\right] ( \nu ,\tau)
=
\vartheta \left[
\begin{array}{c}
a +1 \\
b  \\
\end{array}
\right] (\nu,\tau),
\end{equation}
\begin{equation}\vartheta \left[
\begin{array}{c}
-a \\
0  \\
\end{array}
\right] ( 0 ,\tau)
=
\vartheta \left[
\begin{array}{c}
a \\
0  \\
\end{array}
\right] (0,\tau).
\end{equation}

\begin{figure}[thbp]
 \begin{center}
  \epsfig{file=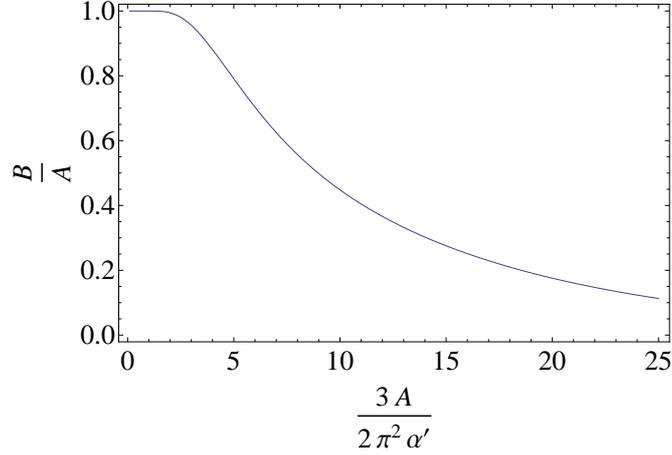,scale=0.6}
 \end{center}
 \caption{$B/A$ vs. $3A/2\pi^2 \alpha'$}
 \label{fig:B/A}
\end{figure}

Then, we see that all of the off-diagonal elements are the same,
\begin{eqnarray}
	c_{01}=c_{12}=c_{20}.
\end{eqnarray}
That is, we can realize the form (\ref{eq:n-mass-33}) 
by explicit calculations.
Figure~\ref{fig:B/A} shows the ratio $B/A=c_{12}/c_{aa}$ in (\ref{eq:n-mass-33}) 
by varying the area $3A/2\pi^2\alpha'$.

\subsection{Phenomenological implication}

Here we discuss phenomenological implication of our result.
The Majorana mass matrix with the form (\ref{eq:n-mass-33}) can be diagonalized by the following matrix,
\begin{eqnarray}\label{eq:mixing}
\left(
\begin{array}{ccc}
\sqrt{2/3}c & 1/\sqrt{3} & -\sqrt{2/3}s \\
   -1/\sqrt{6}c-1/\sqrt{2}s & 1/\sqrt{3} &   1/\sqrt{6}s-1/\sqrt{2}c \\
     -1/\sqrt{6}c+1/\sqrt{2}s & 1/\sqrt{3} &   1/\sqrt{6}s +1/\sqrt{2}c \\
\end{array}
\right),
\end{eqnarray}
where $c=\cos \theta$ and $s= \sin \theta$,
and the eigenvalues are $A-B$, $A+2B$ and $A-B$.
That is, two eigenvalues are degenerate.
This is because the mass matrix (\ref{eq:n-mass-33})  has the additional 
$Z_2$ reflection symmetry $P$ and the symmetry is enhanced into $S_3$.
At any rate, this form of the mixing matrix is interesting, although 
the mass eigenvalues may be not completely realistic.

Suppose that the Dirac neutrino Yukawa couplings and charged lepton mass matrix are 
almost diagonal.\footnote{The $\Delta(27)$ flavor symmetry as well as $\Delta(54)$ flavor symmetry 
may be useful to realize such a form.}
Then, the lepton mixing matrix is obtained as the above matrix (\ref{eq:mixing}).
That is the trimaximal matrix.

When $s=0$, the above matrix becomes the tri-bimaximal mixing matrix.
In field-theoretical model building, the tri-bimaximal mixing matrix can be 
obtained as follows \cite{Altarelli:2010gt,Ishimori:2010au,King:2013eh}.
We start with a larger flavor symmetry and break by vacuum expectation values 
of scalar fields.
However, one assumes that $Z_3$ and $Z_2$ subsymmetries remain in the charged lepton or 
neutrino mass terms.
Then, the tri-bimaximal mixing matrix can be realized.
In our string theory, such a $Z_3$ symmetry is realized by geometrical symmetry 
of the cyclic permutation $Z_3^{C}$, which can not be broken by 
the D-brane instanton effects, although other symmetries are broken.

We may need some corrections to realize the experimental values of 
neutrino masses.\footnote{
To resolve the degeneracy between two mass eigenvalues, it may be important to 
break the $Z_2$ reflection symmetry $P$. 
The full D-brane system, i.e. the full Lagrangian 
of the low-energy effective field theory, may not have such $Z_2$ symmetry and the above 
degeneracy may be resolved by radiative corrections.}
At least, the above results show that we can realize non-trivial mixing 
in the lepton sector even though our assumption above the Dirac masses can not 
be realized.

\section{Conclusion and discussion}

We have studied the flavor structure in 
intersecting D-brane models.
We have discussed the anomalies of flavor symmetries.
Certain symmetries are anomaly-free, 
and anomaly coefficients of discrete symmetries have the specific 
feature.
We have studied the Majorana neutrino masses, 
which can be generated by D-brane instanton effects.
It is found that the mass matrix form with the cyclic permutation symmetry 
can be realized by integrating over the position of 
D-brane instanton.
That would lead to the interesting form of mixing angles.
It is interesting to apply our results for more concrete models.
We would study numerical analyses elsewhere.

In some models, there appear more than one pair of Higgs fields.
Their masses would be generated by D-brane instanton effects.
It would be important to study the form of such Higgs mass matrix.
Also, some of Yukawa couplings may be generated by 
D-brane instanton effects.
Thus, it would be important to extend our analysis to 
Higgs mass matrix and Yukawa matrices.

\subsection*{Acknowledgement}
 The work of Y. H. is supported in part by the 
Grant-in-Aid for  Japan Society for the Promotion of Science (JSPS)
Fellows No.25$\cdot$1107. 
The work of T.K. is supported in part by the Grants-in-Aid for  Scientific
 No. 25400252 from the Ministry of Education, Culture,Sports, Science and 
Technology of Japan.

\appendix

\section{Integration of Wilson line moduli}

Here we integrate Wilson line moduli of the $D2_{M}$-brane. 
Non-zero Wilson line varies the 3 point coupling (\ref{eq:3pcoupling}) to \red{the following form.}

\begin{equation} d_{a}^{i'j'} = C \vartheta \left[
\begin{array}{c}
\frac{a}{I_{cd}} + \frac{i'}{I_{dM}} + \frac{j'}{I_{Mc}}+\frac{\varepsilon}{I_{dM} I_{Mc}} \\
0  \\
\end{array}
\right] \left( \phi ,\frac{iA |I_{cd} I_{dM} I_{Mc}|}{4\pi^{2}\alpha'} \right) .
\label{eq:3pcoupling_WL}
\end{equation}
Here, $\phi$ is Wilson line phase.
The matrix $c_{ab}$ is written by the integration of the position $\varepsilon$ and Wilson line moduli $\phi$.

\begin{eqnarray}c_{ab} & = & \int d\phi \int_{0}^{1} d\varepsilon \vartheta \left[
\begin{array}{c}
-\frac{a}{3} + \varepsilon \\
0  \\
\end{array}
\right] \left( \phi ,\frac{3iA}{4\pi^{2}\alpha'} \right)
\vartheta \left[
\begin{array}{c}
-\frac{b}{3} + \varepsilon \\
0  \\
\end{array}
\right] \left( \phi ,\frac{3iA}{4\pi^{2}\alpha'} \right) \nonumber \\
& = &
\int d\phi \int_{0}^{1} d\varepsilon \sum_{m=1}^{2} 
\vartheta \left[
\begin{array}{c}
-\frac{a}{6} - \frac{b}{6} + \varepsilon + \frac{m}{2}\\
0  \\
\end{array}
\right] \left( 2\phi ,\frac{3iA}{2\pi^{2}\alpha'} \right)  \\
& & \times \vartheta \left[
\begin{array}{c}
-\frac{a}{6} +\frac{b}{6} +\frac{m}{2} \\
0  \\
\end{array}
\right] \left( 0,\frac{3iA}{2\pi^{2}\alpha'} \right). \nonumber
\end{eqnarray}
We get
\begin{eqnarray}
&&\int d\phi \int_{0}^{1} d\epsilon
\vartheta \left[
\begin{array}{c}
-\frac{a}{6} - \frac{b}{6} + \epsilon + \frac{m}{2}\\ \nonumber
0  \\
\end{array}
\right] \left( 2\phi ,2i \frac{3A}{2\pi^{2} \alpha'} \right)\\ \nonumber
& &=
\int d\phi \int_{0}^{1} d\epsilon \sum_{l\in \bf{Z}} e^{- \frac{3A}{\pi \alpha'}\left(-\frac{a}{6}-\frac{b}{6}+\epsilon +\frac{m}{2} +l \right)^{2}+4\pi i \left(-\frac{a}{6}-\frac{b}{6}+\epsilon +\frac{m}{2} + l \right)\phi}\\ \nonumber
& &=
\int d\phi \int_{0}^{1} d\epsilon \sum_{l\in \bf{Z}} e^{- \frac{3A}{\pi \alpha'} \left(-\frac{a}{6}-\frac{b}{6}+\epsilon +\frac{m}{2} + l + i\frac{ 2\pi^{2} \alpha'}{3A}\phi \right)^2 - \frac{4\pi^{3} \alpha' \phi^{2}}{3A}}\\
& &=
\sqrt{\frac{\pi^{2} \alpha'}{3A}}\int d\phi e^{- \frac{4\pi^{3} \alpha' \phi^{2}}{3A}}.
\end{eqnarray}
\red{This factor is independent of flavor index, but universal.
Thus, the integration of Wilson line moduli does not affect flavor structure.}



\begin{thebibliography}{99}


\bibitem{Altarelli:2010gt} 
  G.~Altarelli and F.~Feruglio,
  Rev.\ Mod.\ Phys.\  {\bf 82}, 2701 (2010)
  [arXiv:1002.0211 [hep-ph]].


\bibitem{Ishimori:2010au} 
  H.~Ishimori, T.~Kobayashi, H.~Ohki, Y.~Shimizu, H.~Okada and M.~Tanimoto,
  Prog.\ Theor.\ Phys.\ Suppl.\  {\bf 183}, 1 (2010)
  [arXiv:1003.3552 [hep-th]]; 
  %
  Lect.\ Notes Phys.\  {\bf 858}, pp.1 (2012);
%
  Fortsch.\ Phys.\  {\bf 61}, 441 (2013).


\bibitem{King:2013eh} 
  S.~F.~King and C.~Luhn,
  Rept.\ Prog.\ Phys.\  {\bf 76}, 056201 (2013)
  [arXiv:1301.1340 [hep-ph]].

\bibitem{Harrison:2002er} 
  P.~F.~Harrison, D.~H.~Perkins and W.~G.~Scott,
  Phys.\ Lett.\ B {\bf 530}, 167 (2002)
  [hep-ph/0202074];
 P.~F.~Harrison and W.~G.~Scott,
  Phys.\ Lett.\ B {\bf 535}, 163 (2002)
  [hep-ph/0203209];
  Phys.\ Lett.\ B {\bf 557}, 76 (2003)
  [hep-ph/0302025].

\bibitem{Ibanez}
  L.~E.~Ibanez and A.~M.~Uranga,
  ``String theory and particle physics: 
  An introduction to string phenomenology,''
  Cambridge University Press (2012).


\bibitem{Bachas:1995ik} 
  C.~Bachas,
hep-th/9503030.  


\bibitem{Berkooz:1996km} 
  M.~Berkooz, M.~R.~Douglas and R.~G.~Leigh,
Nucl.\ Phys.\ B {\bf 480}, 265 (1996)  [hep-th/9606139].  

\bibitem{Blumenhagen:2000wh} 
  R.~Blumenhagen, L.~Goerlich, B.~Kors and D.~Lust,
JHEP {\bf 0010}, 006 (2000)  [hep-th/0007024].  


\bibitem{Aldazabal:2000dg} 
  G.~Aldazabal, S.~Franco, L.~E.~Ibanez, R.~Rabadan and A.~M.~Uranga,
  J.\ Math.\ Phys.\  {\bf 42}, 3103 (2001)
  [hep-th/0011073];
  JHEP {\bf 0102}, 047 (2001)
  [hep-ph/0011132].



\bibitem{Angelantonj:2000hi} 
  C.~Angelantonj, I.~Antoniadis, E.~Dudas and A.~Sagnotti,
Phys.\ Lett.\ B {\bf 489}, 223 (2000)  [hep-th/0007090].  





\bibitem{Ibanez:2001nd} 
  L.~E.~Ibanez, F.~Marchesano and R.~Rabadan,
JHEP {\bf 0111}, 002 (2001)  [hep-th/0105155].  




\bibitem{Blumenhagen:2006ci} 
  R.~Blumenhagen, B.~Kors, D.~Lust and S.~Stieberger,
Phys.\ Rept.\  {\bf 445}, 1 (2007)  [hep-th/0610327].  


\bibitem{Abe:2009vi}
  H.~Abe, K.~-S.~Choi, T.~Kobayashi and H.~Ohki,
  Nucl.\ Phys.\ B {\bf 820} (2009) 317
  [arXiv:0904.2631 [hep-ph]].
%
\bibitem{Abe:2009uz}
   H.~Abe, K.~-S.~Choi, T.~Kobayashi and H.~Ohki,
  Phys.\ Rev.\ D {\bf 80} (2009) 126006
  [arXiv:0907.5274 [hep-th]]; 
%
  Phys.\ Rev.\ D {\bf 81} (2010) 126003
  [arXiv:1001.1788 [hep-th]].


\bibitem{BerasaluceGonzalez:2012vb} 
  M.~Berasaluce-Gonzalez, P.~G.~Camara, F.~Marchesano, D.~Regalado and A.~M.~Uranga,
  JHEP {\bf 1209}, 059 (2012)
  [arXiv:1206.2383 [hep-th]];
  F.~Marchesano, D.~Regalado and L.~Vazquez-Mercado,
  JHEP {\bf 1309}, 028 (2013)
  [arXiv:1306.1284 [hep-th]].


\bibitem{Kobayashi:2004ya}
  T.~Kobayashi, S.~Raby and R.~J.~Zhang,
  Nucl.\ Phys.\  B {\bf 704}, 3 (2005)
  [arXiv:hep-ph/0409098];
%
  T.~Kobayashi, H.~P.~Nilles, F.~Ploger, S.~Raby and M.~Ratz,
  Nucl.\ Phys.\  B {\bf 768}, 135 (2007)
  [arXiv:hep-ph/0611020];
%
  P.~Ko, T.~Kobayashi, J.~h.~Park and S.~Raby,
  Phys.\ Rev.\  D {\bf 76}, 035005 (2007)
  [Erratum-ibid.\  D {\bf 76}, 059901 (2007)]
  [arXiv:0704.2807 [hep-ph]].

\bibitem{Higaki:2005ie} 
  T.~Higaki, N.~Kitazawa, T.~Kobayashi and K.~-j.~Takahashi,
  Phys.\ Rev.\ D {\bf 72}, 086003 (2005)
  [hep-th/0504019].


\bibitem{Hamidi:1986vh} 
  S.~Hamidi and C.~Vafa,
  Nucl.\ Phys.\ B {\bf 279}, 465 (1987);
  L.~J.~Dixon, D.~Friedan, E.~J.~Martinec and S.~H.~Shenker,
  Nucl.\ Phys.\ B {\bf 282}, 13 (1987);
%
  T.~T.~Burwick, R.~K.~Kaiser and H.~F.~Muller,
  Nucl.\ Phys.\ B {\bf 355}, 689 (1991);
%
  J.~Erler, D.~Jungnickel, M.~Spalinski and S.~Stieberger,
  Nucl.\ Phys.\ B {\bf 397}, 379 (1993)
  [hep-th/9207049];
%
  K.~-S.~Choi and T.~Kobayashi,
  Nucl.\ Phys.\ B {\bf 797}, 295 (2008)
  [arXiv:0711.4894 [hep-th]].
%
  T.~Kobayashi, S.~L.~Parameswaran, S.~Ramos-Sanchez and I.~Zavala,
  JHEP {\bf 1205}, 008 (2012)
  [Erratum-ibid.\  {\bf 1212}, 049 (2012)]
  [arXiv:1107.2137 [hep-th]].
  
  
  
\bibitem{Cvetic:2003ch} 
  M.~Cvetic and I.~Papadimitriou,
  Phys.\ Rev.\ D {\bf 68}, 046001 (2003)
  [Erratum-ibid.\ D {\bf 70}, 029903 (2004)]
  [hep-th/0303083];
  S.~A.~Abel and A.~W.~Owen,
  Nucl.\ Phys.\ B {\bf 663}, 197 (2003)
  [hep-th/0303124];
%
  D.~Cremades, L.~E.~Ibanez and F.~Marchesano,
  JHEP {\bf 0307}, 038 (2003)
  [hep-th/0302105];
  S.~A.~Abel and A.~W.~Owen,
  Nucl.\ Phys.\ B {\bf 682}, 183 (2004)
  [hep-th/0310257].

\bibitem{Cremades:2004wa} 
  D.~Cremades, L.~E.~Ibanez and F.~Marchesano,
  JHEP {\bf 0405}, 079 (2004)
  [hep-th/0404229];
%
  H.~Abe, K.~-S.~Choi, T.~Kobayashi and H.~Ohki,
  JHEP {\bf 0906}, 080 (2009)
  [arXiv:0903.3800 [hep-th]];
  Y.~Hamada and T.~Kobayashi,
  Prog.\ Theor.\ Phys.\  {\bf 128},  903 (2012)
  [arXiv:1207.6867 [hep-th]].


\bibitem{BerasaluceGonzalez:2011wy} 
  M.~Berasaluce-Gonzalez, L.~E.~Ibanez, P.~Soler and A.~M.~Uranga,
  JHEP {\bf 1112}, 113 (2011)
  [arXiv:1106.4169 [hep-th]].
  
\bibitem{Ibanez:2012wg} 
  L.~E.~Ibanez, A.~N.~Schellekens and A.~M.~Uranga,
  Nucl.\ Phys.\ B {\bf 865}, 509 (2012)
  [arXiv:1205.5364 [hep-th]].

\bibitem{BerasaluceGonzalez:2012zn} 
  M.~Berasaluce-Gonzalez, P.~G.~Camara, F.~Marchesano and A.~M.~Uranga,
  JHEP {\bf 1304}, 138 (2013)
  [arXiv:1211.5317 [hep-th]].
  
\bibitem{Anastasopoulos:2012zu} 
  P.~Anastasopoulos, M.~Cvetic, R.~Richter and P.~K.~S.~Vaudrevange,
  JHEP {\bf 1303}, 011 (2013)
  [arXiv:1211.1017 [hep-th]].
  
 
\bibitem{Honecker:2013hda} 
  G.~Honecker and W.~Staessens,
  JHEP {\bf 1310}, 146 (2013)
  [arXiv:1303.4415 [hep-th]].
 
\bibitem{Bizet:2013gf} 
  N.~G.~Cabo Bizet, T.~Kobayashi, D.~K.~Mayorga Pena, S.~L.~Parameswaran, M.~Schmitz and I.~Zavala,
  JHEP {\bf 1305}, 076 (2013)
  [arXiv:1301.2322 [hep-th]].
  
\bibitem{Nilles:2013lda} 
  H.~P.~Nilles, S.~ulRamos-Sanchez, M.~Ratz and P.~K.~S.~Vaudrevange,
  Phys.\ Lett.\ B {\bf 726}, 876 (2013)
  [arXiv:1308.3435 [hep-th]].

\bibitem{Bizet:2013wha} 
  N.~G.~C.~Bizet, T.~Kobayashi, D.~K.~M.~Pena, S.~L.~Parameswaran, M.~Schmitz and I.~Zavala,
  arXiv:1308.5669 [hep-th].

  
\bibitem{Araki:2008ek} 
  T.~Araki, T.~Kobayashi, J.~Kubo, S.~Ramos-Sanchez, M.~Ratz and P.~K.~S.~Vaudrevange,
  Nucl.\ Phys.\ B {\bf 805}, 124 (2008)
  [arXiv:0805.0207 [hep-th]].
  
\bibitem{Araki:2007ss} 
  T.~Araki, K.~-S.~Choi, T.~Kobayashi, J.~Kubo and H.~Ohki,
  Phys.\ Rev.\ D {\bf 76}, 066006 (2007)
  [arXiv:0705.3075 [hep-ph]].


\bibitem{Blumenhagen:2006xt} 
  R.~Blumenhagen, M.~Cvetic and T.~Weigand,
  Nucl.\ Phys.\ B {\bf 771}, 113 (2007)
  [hep-th/0609191].

\bibitem{Blumenhagen:2009qh} 
  R.~Blumenhagen, M.~Cvetic, S.~Kachru and T.~Weigand,
  Ann.\ Rev.\ Nucl.\ Part.\ Sci.\  {\bf 59}, 269 (2009)
  [arXiv:0902.3251 [hep-th]].


\bibitem{Ibanez:2006da} 
  L.~E.~Ibanez and A.~M.~Uranga,
JHEP {\bf 0703}, 052 (2007)  [hep-th/0609213].


\bibitem{Ibanez:2007rs} 
  L.~E.~Ibanez, A.~N.~Schellekens and A.~M.~Uranga,
  JHEP {\bf 0706}, 011 (2007)
  [arXiv:0704.1079 [hep-th]];
  S.~Antusch, L.~E.~Ibanez and T.~Macri,
  JHEP {\bf 0709}, 087 (2007)
  [arXiv:0706.2132 [hep-ph]].



\bibitem{Cvetic:2007ku} 
  M.~Cvetic, R.~Richter and T.~Weigand,
  Phys.\ Rev.\ D {\bf 76}, 086002 (2007)
  [hep-th/0703028].




\bibitem{Cvetic:2001nr} 
  M.~Cvetic, G.~Shiu and A.~M.~Uranga,
  Nucl.\ Phys.\ B {\bf 615}, 3 (2001)
  [hep-th/0107166].






\bibitem{Ibanez:1991hv} 
  L.~E.~Ibanez and G.~G.~Ross,
  Phys.\ Lett.\ B {\bf 260}, 291 (1991).
  
\bibitem{Banks:1991xj} 
  T.~Banks and M.~Dine,
  Phys.\ Rev.\ D {\bf 45}, 1424 (1992)
  [hep-th/9109045].




\end{thebibliography}
\end{document}